\documentstyle[12pt]{article}
\renewcommand{\thefootnote}{\fnsymbol{footnote}}
\begin{document}
\begin{flushright}
Frankfurt preprint UFTP--406\\
Columbia preprint CU--TP--733
\end{flushright}
\vspace*{1cm}
\setcounter{footnote}{1}
\begin{center}
{\Large\bf Relativistic Hydrodynamics for Heavy--Ion Collisions:}\\
{\Large\bf Freeze--Out and Particle Spectra\footnote{Supported by 
BMBF, DFG, GSI, and DOE.}}
\\[1cm]
Stefan Bernard, Joachim A.\ Maruhn, Walter Greiner \\ ~~ \\
{\small Institut f\"ur Theoretische Physik der J.W.\ Goethe--Universit\"at} \\
{\small Robert--Mayer--Str.\ 10, D--60054 Frankfurt/M., Germany}
\\ ~~ \\
Dirk H.\ Rischke\footnote{Partially supported by the
Alexander von Humboldt--Stiftung under
the Feodor--Lynen program and by the Director, Office of Energy
Research, Division of Nuclear Physics of the Office of 
High Energy and Nuclear Physics of the U.S. Department of 
Energy under Contract No.\ DE-FG-02-93ER-40764.} \\ ~~ \\
{\small Physics Department, Pupin
Physics Laboratories, Columbia University} \\
{\small 538 W 120th Street, New
York, NY 10027, U.S.A.}
\\ ~~ \\ ~~ \\
{\large January 1996}
\\[1cm]
\end{center}
\begin{abstract}
We investigate freeze--out in hydrodynamic models for
relativistic heavy--ion collisions. In particular, 
instantaneous freeze--out across a hypersurface of constant
temperature (``isothermal'' freeze--out) is compared with that across
a hypersurface at constant time in the center-of-momentum frame
(``isochronous'' freeze--out). For one--dimensional (longitudinal)
expansion the rapidity distributions
are shown to differ significantly in the two scenarios, while the
transverse momentum spectra are remarkably similar. We also investigate
the rapidity distribution in greater detail and show that
the Gaussian-like shape of this distribution commonly associated
with the Landau expansion model in general emerges only if one neglects
contributions from time-like parts of the isothermal 
freeze--out hypersurface.
\end{abstract}
\renewcommand{\thefootnote}{\arabic{footnote}}
\setcounter{footnote}{0}
\newpage

\section{Introduction}

Relativistic hydrodynamic models are frequently applied to
study high--energy heavy--ion collisions \cite{Stoecker,Strottman},
because they simply represent (local) energy--momentum and charge 
conservation. Their main advantage is that they provide a {\em direct\/}
link between space--time dynamics, i.e., observable flow phenomena, and
the nuclear matter equation of state. They are therefore a unique tool
to investigate e.g.\ effects of phase transitions, such as the
deconfinement transition to the quark--gluon plasma,
on the dynamics of the reaction
\cite{Strottman,Csernai,flow,lifetime}.

Relativistic hydrodynamics requires the assumption of
{\em local thermodynamical equi\-li\-bri\-um}\footnote{A well-defined theory
of relativistic dissipative hydrodynamics,
which would allow for deviations from local thermodynamical equilibrium,
does not yet exist, see for instance the discussion in \cite{Strottman}.},
i.e., it describes the evolution of an {\em ideal\/} fluid \cite{LL6}.
This assumption obviously breaks down in the
very early stages of ultrarelativistic heavy--ion collisions
(the stopping power of nuclear matter is insufficient to produce
immediate thermalization of matter in the collision zone) and in
the very late stages of heavy--ion collisions at all energies.
In this stage, the interaction rate between individual particles becomes
too small, or equivalently, their mean free path between collisions
becomes too long, to maintain local thermodynamical equi\-li\-bri\-um
in the expanding system, and non--equilibrium processes become dominant.
Eventually, particles will decouple completely
and stream freely until they reach the detector.

In order to calculate experimental observables
it is essential to incorporate this so-called {\em freeze--out\/}
into the hydrodynamical simulation of a heavy--ion collision.
The long mean free path or the small interaction rate,
respectively, suggest a treatment of the freeze--out via kinetic theory
\cite{LL10}. Up to now, however, no attempt has been made to combine
the hydrodynamical and the kinetic description to model the late expansion
stage of a heavy--ion collision. The most simple and therefore often
used method is to assume an {\em instantaneous\/} transition from
fluid-dynamical motion to free--streaming. Note that microscopic
models for heavy--ion collisions \cite{MicroModel} do not encounter
this problem, since they follow the trajectories of individual particles.
On the other hand, their disadvantage is that they do not provide a
direct link to the nuclear matter equation of state and, when
based on binary scatterings, become invalid for higher densities.

In this paper we present a systematic comparison of two different
instantaneous freeze--out scenarios presently used in the literature,
namely the freeze--out across a space--time hypersurface
defined by a constant temperature (referred to as ``isothermal''
freeze--out in the following) and that across one defined by a
constant time in the center-of-momentum (CM) frame of the system,
respectively (subsequently termed ``isochronous'' freeze--out).
Contributions from time-like and space-like parts of the
hypersurface\footnote{A definition
of time-like and space-like surfaces is given in Section 3.} are
discussed separately and it is shown that the often used approximation
of neglecting the time-like contributions in the isothermal
freeze--out scenario leads to
serious distortions of the rapidity spectra,
and moreover violates the conservation laws.

The above freeze--out scenarios are not strictly self-consistent,
because the hypersurface is determined {\em after\/} the
fluid-dynamical problem has been solved in the {\em whole\/} forward
light cone. In other words, the influence of the freeze--out of particles
from the (time-like or space-like) surface of the fluid on the
dynamical evolution is neglected.
Although attempts to treat this problem analytically
have been made \cite{Sinyukov}, up to now no
generally accepted solution exists.

This paper is organized as follows. In Section 2 we present the
underlying relativistic hydrodynamic expansion model.
In Section 3 we introduce the two different freeze--out scenarios.
Section 4 contains a discussion of single--inclusive rapidity distributions
and transverse momentum spectra. In Section 5 we present our results
and close with a summary in Section 6. We use natural units, $\hbar=k_B=c=1$.

\section{The expansion model}

Since we want to focus on the effects of different
freeze--out scenarios on the final particle spectra, we
choose a particularly simple and transparent
hydrodynamic expansion model, namely
the one--dimensional expansion of a finite slab of matter with an
ultrarelativistic ideal gas equation of state.
This model was first studied by Landau et al.\
\cite{Landau}.
The expansion with Landau initial conditions
can to first approximation serve to describe the
evolution of (net) baryon--free matter
in the central region of ultrarelativistic
heavy--ion collisions $(\sqrt{s} \sim 100\,\rm{AGeV})$.
In the following we briefly outline the essential ingredients and
results of the Landau expansion model.

In general, relativistic hydrodynamics implies (local) energy--momentum
conservation,
\begin{equation} \label{dtmn}
\partial_{\mu} T^{\mu \nu} =0 ~,
\end{equation}
where for an ideal fluid the energy--momentum tensor reads
\begin{equation} \label{tmn}
T^{\mu \nu} = (\epsilon +p) u^{\mu} u^{\nu} - p g^{\mu \nu} ~.
\end{equation}
Here, $\epsilon$ and $p$ denote the energy density and pressure in the local
rest frame of matter, $g^{\mu \nu}={\rm diag}(+1,-1,-1,-1)$ is the metric
tensor, and $u^{\mu}$ is the fluid 4--velocity
($u^{\mu} u_{\mu}=1$).
The system of eqs.\ (\ref{dtmn}, \ref{tmn})
is closed by choosing an equation of state for the matter under
consideration. For an ultrarelativistic ideal gas this is
\begin{equation} \label{EoS}
p=c_s^2\, \epsilon~,
\end{equation}
with $c_s=1/\sqrt{3}$ being the velocity of sound.
For the particular application to heavy--ion collisions,
we may think of an ideal (massless) pion gas.
Then, the energy density is given in terms of the temperature $T$ as
\begin{equation} \label{epi}
\epsilon=g\, \frac{{\pi}^{2}}{30}\,T^4 ~,
\end{equation}
where $g=3$ is the degeneracy factor for pions (assuming isospin
symmetry). The particle density $n$ in the local rest frame is
\begin{equation} \label{npi}
n=g\, \frac{\zeta(3)}{\pi^2}\, T^3 ~.
\end{equation}

In the CM frame, the initial conditions for 
the one--dimensional expansion of a finite slab of length $2\, L$ into 
vacuum \cite{Landau} are
\begin{eqnarray} \label{inie}
\epsilon (0,z) &=& \left\{ \begin{array}{ll}
                \epsilon_0 &,~~|z| \leq L \\
                         0 &,~~|z| > L ~,
                           \end{array} \right. \\
v(0,z) &=& \left\{ \begin{array}{ll}
                        -1 &,~~-\infty < z < -L \\
                         0 &,~~-L \leq z \leq L \\
                         1 &,~~L < z < \infty ~.
                           \end{array} \right.
\end{eqnarray}
Note that all lengths and times scale with $L$.

The system evolves as follows, cf.\ Fig.\ 1. Initially, the discontinuities
at $z= \pm L$ decay into two simple (Riemann)
rarefaction waves. The wave fronts move into the slab with
velocities $\pm c_s$ and into the vacuum with
the speed of light $\pm 1$. At the CM time $t_m=L/c_s$
the two simple waves meet at the center
and begin to overlap. At later times $t>t_m$
there are three distinct regions of hydrodynamic flow in the
forward light cone.
We call the one at the center of the system,
where the two rarefaction waves overlap,
the {\em Landau region\/}
\cite{Landau,Achenbach,Srivastava}, because Landau was the first
to determine the hydrodynamic solution in this region.
The other two are called {\em Riemann regions},
because here the hydrodynamic solution is still given by the original
simple Riemann rarefaction waves.

Since we ultimately want to apply the freeze--out scenarios discussed
in this work to more complicated, three--dimensional simulations,
we study the Landau expansion numerically.
For the numerical solution we employ the explicit SHASTA algorithm
\cite{BorisBook}. The applicability of this version of SHASTA to
relativistic one--dimensional compression and expansion problems
in thermodynamically normal matter has been
demonstrated in Refs.\ \cite{RischkeBer,Schneider}.
We employ a reduced antidiffusion and 
also a simple full--step method for updating
source terms\footnote{The half--step treatment of source terms
is necessary only to appropriately describe compression
shocks \cite{RischkeBer}.}.

The initial temperature profile and typical profiles for times $t>t_m$ 
are shown in Fig.\ 2. The temperature
is calculated from the local energy density
$\epsilon$ according to (\ref{epi}), and normalized to the initial
temperature $T_{0}$.
Note that the overlapping rarefaction waves in the Landau region
produce a local minimum of energy density and temperature
at fixed CM times $t>t_{m}$. This is essentially a relativistic
effect: faster matter at larger $|z|$ experiences more time dilation
and thus cools not as fast in the CM frame as
slower matter near the center \cite{RischkeBer}.

\section{The freeze--out scenarios}

\subsection{General aspects}

A particle ``freezes out'' when its interactions with the
rest of the system cease. In principle, this can happen at any
space--time point of the many--particle evolution,
depending on the (finite) mean free path of the particular particle.
In a one--component system, the mean free path is estimated as
\begin{equation} \label{mfp}
\lambda \simeq \frac{1}{n \sigma}~,
\end{equation}
where $\sigma$ is the (total) cross section and $n$ the (local)
comoving particle density. If $\lambda$ is
small enough such that local thermodynamical equilibrium is
established, ideal hydrodynamics is valid (and $n$ is, for a pionic
system, given by eq.\ (\ref{npi})). For larger $\lambda$
non--equilibrium effects become increasingly important, and if
$\lambda$ exceeds the system's dimension $D$, the
latter starts to decouple into free--streaming particles.
To describe the system's evolution in
the latter two stages in principle requires kinetic theory.

Up to now, drastic simplifications of the freeze--out stage
were made for practical calculations
\cite{Landau,CooperFrye}.
The first assumption is that the intermediate region where $\lambda \sim D$
has negligible extent and can be approximated by a hypersurface in 
space--time.
When a fluid element crosses this hypersurface, particles contained
in that element freeze out instantaneously.
The second assumption is that the freeze--out of particles does not
change the hydrodynamical evolution.
Because of energy and momentum carried away by the leaving 
particles, the hydrodynamical solution
is obviously influenced. Also, the decoupling of a particle
from the flow perturbes (local) thermodynamic
and chemical equilibrium in the respective fluid element. In general,
the freeze--out should be treated in a self-consistent way taking
all these effects into account. This, however, is out of the scope of
the present work.

As mentioned in the Introduction, here we also assume instantaneous
freeze--out across a hypersurface. We focus on two different scenarios.
The first, referred to in the following as
{\em isothermal freeze--out}, corresponds to freeze--out
at a constant fluid temperature $T_f$.
The second corresponds to freeze--out at a constant
CM time $t_f$ and is referred to as {\em isochronous freeze--out}.

\subsection{Isothermal freeze--out}

The isothermal freeze--out is based on assumptions originally
formulated by Landau \cite{Landau}.
He argued that the freeze--out of an interacting hadronic fluid into
free--streaming particles takes place at a certain freeze--out
temperature $T_{f}$ which, according to eq.\ (\ref{mfp}) with $\lambda
\sim D$, depends only weakly on the system's size, $T_f \sim D^{-1/3}$.
Thus, the isotherm with $T=T_f$ defines the freeze--out hypersurface.
For pions we have (roughly) $\sigma \sim 1$ fm$^2$ (at the
energy scale relevant for pion--pion collisions in a thermal ensemble).
We also assume $D \sim 2\, t_f \simeq 20\, L \sim 10$ fm (cf.\ Fig.\ 2), 
where the initial
size of the hot central region was estimated to be $2\, L \sim 2\, t_0
\sim 1$ fm (for an equilibration time $t_0 \sim 0.5$ fm after impact),
and obtain with eq.\ (\ref{npi}) the freeze--out
criterion $T_{f} \simeq m_{\pi}$, where $m_{\pi}\simeq 138$ MeV
is the pion mass.

In order to apply this scenario to our underlying expansion model
presented in Section 2, the isotherm of temperature $T_{f}$
has to be found numerically on the calculational grid.
As an example, Fig.\ 1 shows the isotherm $T_f=0.4\, T_0$.
The isotherm is in very good agreement with that
calculated with the (semi--analytical) method of
characteristics, see
\cite{BaymFri}.

A hypersurface element is called space-like if its normal vector is time-like,
and vice versa. The whole isothermal hypersurface can be divided into
space- and time-like parts according to Fig.\ 1.

\subsection{Isochronous freeze--out}

Due to its simplicity, the isochronous freeze--out scenario is widely 
used in (3+1)--dimensional
one-- and multi--fluid-dynamics \cite{Stoecker}. The freeze--out condition 
one commonly assumes is that the {\em average\/}
temperature of the system at CM time $t_f$ drops below a given value $T_f$,
\begin{equation} \label{aTglob}
\bar{T} (t_{f})=T_{f} ~,
\end{equation}
where $\bar{T} (t) $ is calculated according to
\begin{equation} \label{average}
\bar{T}(t)=\frac{ \int_{-L-t}^{L+t}{\rm d}z\, T \gamma n }
         { \int_{-L-t}^{L+t}{\rm d}z\, \gamma n }~,
\end{equation}
where $\gamma=(1-v^2)^{-1/2}$ and $n$ is the local rest frame density
given by eq.\ (\ref{npi}).

Note that the hypersurface for isochronous freeze--out is purely
space-like. Fig.\ 1 shows the hypersurfaces at CM times $t_f$ when
(a) the {\em average\/} temperature is $\bar{T}=0.4\, T_0$
($t \simeq 9.2\,\rm{L}$, lower line), and (b)
when the {\em maximum\/} temperature in the system
has dropped below $T_f$ (``late'' isochronous freeze--out,
$t \simeq 12.6\,\rm{L}$, upper line).
Fig.\ 2 shows the corresponding temperature profiles.

Although the isochronous freeze--out surfaces are purely space-like,
we distinguish ``time-like'' (dotted) and ``space-like''
(full line) parts as shown in Fig.\ 1. This is motivated
by the situation at ``late'' isochronous freeze--out:
the energy and momentum
associated with particles freezing out across $\Sigma_{t}$
in the isothermal freeze--out scenario later has
to cross the dotted, ``time-like'' parts
of the ``late'' isochronous freeze--out surface, because it cannot
leave the forward light cone. On the other hand, the energy and momentum
freezing out across $\Sigma_{s}$ is the same
that freezes out across the  ``space-like'' part of the ``late''
isochronous freeze--out surface.

\section{Calculation of the spectra}

\subsection{The Cooper--Frye formula}

In both freeze--out scenarios, the
distribution of particles that have decoupled from the fluid is determined
as follows. The (Lorentz--invariant) momentum--space distribution
of particles crossing a hypersurface $\Sigma$ in Minkowski space
is given by \cite{CooperFrye}
\begin{equation} \label{Coo1}
  E\, \frac{{\rm d} N}{{\rm d}^{3} {\bf k}} =
    \int_{\Sigma} {\rm d}{\sigma}_{\mu}k^{\mu}\, f(k^{\nu}u_{\nu})~,
\end{equation}
where $f(k^{\nu}u_{\nu})$ is the local equilibrium distribution function
\begin{equation} \label{bose}
  f(k^{\nu}u_{\nu}) =  \frac{g}{(2\pi)^3}\frac{1}{\exp (k^{\nu}u_{\nu}/T)-1}~,
\end{equation}
and ${\rm d}\sigma_{\mu}$ is the normal vector on
an (infinitesimal) element of the hypersurface $\Sigma$.
${\rm d} \sigma_{\mu}$ is naturally chosen to point outwards
with respect to the hotter interior of $\Sigma$, since
(\ref{Coo1}) is supposed to give the momentum distribution of
particles decoupling from the fluid.
In our case of one--dimensional geometry,
the transverse dimensions enter only as
a (constant) transverse area factor $A$.
The parametric integration over the hypersurface $\Sigma$ in (\ref{Coo1})
is performed by distinguishing space-like parts $\Sigma_s$
(with $z$ as integration variable) and time-like parts
$\Sigma_t$ (with $t$ as integration variable), cf.\ Fig.\ 1. This yields the
total momentum--space distribution
\begin{eqnarray} \label{hypercooper}
  \frac{{\rm d}N}{{\rm d} y k_{\bot}{\rm d}k_{\bot}} & = & \frac{gA}{4{\pi}^2}
  \,m_{\bot}
  \left[
  \int_{\Sigma_s} {\rm d}z~
  \frac{\sinh y (\partial t/\partial z)_{\Sigma}-\cosh y}
                                {\exp \{ m_{\bot}\cosh({\alpha}-y)/T
  \} -1} \right. \nonumber \\
 &   & \left.
  +\int_{\Sigma_t} {\rm d} t~
  \frac{\sinh y-\cosh y
  (\partial z/\partial t)_{\Sigma}}
     {\exp \{ m_{\bot}\cosh({\alpha}-y)/T \} -1} \right]~,\label{cp3}
\end{eqnarray}
where $ y = {\rm Artanh} (k_{\|}/E)$ is the
(longitudinal) particle rapidity,
$m_{\bot}=({\bf k}^2_{\bot} + m^2)^{1/2}$ is the transverse
mass (which reduces to $k_{\bot}$ for massless particles), and
$\alpha = {\rm Artanh} (v)$ is the (longitudinal) fluid rapidity.
$(\partial t/\partial z)_{\Sigma}$ and $(\partial z/\partial t)_{\Sigma}$ 
are the (local) slope of the space-like and the inverse slope of the time-like 
hypersurface element, respectively.
Note that the parametric integration in eq.\ (\ref{hypercooper}) holds
for a mathematically positive orientation of $\Sigma$.

\subsection{Application to isothermal freeze--out}

We now apply eq.\ (\ref{cp3}) to the isothermal freeze--out
scenario. Along the isotherm, $T=T_f={\rm const.}$ and only $\alpha$
depends on position or time, respectively.
The rapidity distribution is obtained by integrating (\ref{cp3})
over transverse momentum. For massless particles,
\begin{eqnarray}
  \frac{1}{A{T_0}^3}\frac{{\rm d}N}{{\rm d}y} & = & \frac{g\zeta(3)}{2\pi^2}
                                      \left( \frac{T_f}{T_0} \right)^3
\left[  \int_{\Sigma_s} {\rm d}z~\frac{
    \sinh y (\partial t/\partial z)_{\Sigma}-\cosh y }{
     \cosh^3({\alpha}(z)-y)} \right. \nonumber  \\
 &   & \left. + \int_{\Sigma_t} {\rm d} t~\frac{
    \sinh y-\cosh y (\partial z/\partial t)_{\Sigma} }{
    \cosh^3({\alpha}(t)-y)} \right]~.\label{cp9}
\end{eqnarray}
Analogously, the transverse momentum distribution is obtained integrating
over $y$,
\begin{eqnarray}
  \frac{1}{AT_0} \frac{{\rm d}N}{k_{\bot}{\rm d}k_{\bot}} & = &
  \frac{g}{4 \pi^2}~\frac{k_{\bot}}{T_0} ~ \left[
  \int_{\Sigma_s} {\rm d}z~\int_{-\infty}^{\infty}
   {\rm d}y~\frac{\sinh y (\partial t/\partial z)_{\Sigma}-\cosh y}{\exp
 \{ (k_{\bot}/T_0)(T_0/T_f) \cosh({\alpha(z)}-y) \}-1} \right. \nonumber \\
 &   & \left. +  \int_{\Sigma_t} {\rm d}t~\int_{-\infty}^{\infty}
   {\rm d}y~\frac{\sinh y-\cosh y (\partial z/\partial t)_{\Sigma}}{\exp
 \{ (k_{\bot}/T_0)(T_0/T_f) \cosh({\alpha(t)}-y) \}-1} \right]~.
  \label{cp11}
\end{eqnarray}

\subsection{Application to isochronous freeze--out}

In the isochronous freeze--out scenario the hypersurface $\Sigma$ has
no time-like part and the second term in (\ref{cp3}) vanishes. Moreover,
also the first term in the numerator of the remaining term vanishes
due to $\partial t/\partial z = 0$ for this particular hypersurface.
The temperature along the hypersurface, however, is no longer constant.
Thus the rapidity distribution becomes
\begin{equation} \label{cp13}
  \frac{1}{A T_0^3}\frac{{\rm d}N}{{\rm d}y} = \frac{g\zeta(3)}{2\pi^2}
  \int_{-L-t_f}^{L+t_f} {\rm d}z \left( \frac{T(z)}{T_0} \right)^3 \frac{
  \cosh y }{\cosh^3({\alpha}(z)-y)}~,
\end{equation}
while the transverse momentum distribution is
\begin{equation} \label{cp14}
  \frac{1}{AT_0} \frac{{\rm d}N}{k_{\bot}{\rm d}k_{\bot}} = \frac{g}{4\pi^2}
  ~\frac{k_{\bot}}{T_0}~ \int_{-L-t_f}^{L+t_f} {\rm d}z~
  \int_{-\infty}^{\infty} {\rm d} y~\frac{\cosh y}{\exp
 \{ (k_{\bot}/T_0)(T_0/T(z)) \cosh({\alpha(z)}-y) \} -1}~.
\end{equation}
Note that the boundaries of the $z$--integration correspond
to the light cone.

\section{Results}

In this section we discuss the pion rapidity and transverse momentum
spectra for isothermal and isochronous freeze--out using the
expansion model of Section 2. Note that for ideal fluid dynamics,
entropy is conserved. Since for a gas of massless particles the
particle number density is proportional to the
entropy density, $n= 45\, \zeta(3)s/2\,\pi^4$, also the total
particle number is conserved.
The rapidity distributions presented in this section are numerically
checked to fulfill this conservation law to better than $2\%$ accuracy
(across hypersurfaces bounded by the light cone).

In Fig.\ 3 we present the
rapidity spectrum for isothermal freeze--out, eq.\ (\ref{cp9}),
along the isotherm $T_f=0.4\, T_0$ that was shown in Fig.\ 1.
The contributions from the different parts of
the hypersurface as indicated in Fig.\ 1 are shown separately.
The total spectrum (full line) shows a double--peak
structure, and a small local maximum at $y=0$.
The time-like contributions (dashed) can become negative for particles with
a longitudinal velocity $\tanh y$ which is smaller than the inverse 
slope of the time-like hypersurface, because then
$\sinh y - \cosh y (\partial z/\partial t)_{\Sigma} = \cosh y [ \tanh y
- (\partial z/\partial t)_{\Sigma}] <0$.
These contributions correspond to particles that are overtaken
by the freeze--out condition, and thus in principle
re-enter the fluid instead of freezing out.
Note that it is important to count these particles with their correct
(negative) sign in order to ensure energy--momentum, entropy, and
particle number conservation.
As can be seen in Fig.\ 3, the space-like contributions to the
spectrum (dash-dotted) are always positive, due to 
$|(\partial t / \partial z)_{\Sigma}| < 1$.

The transverse momentum spectrum, eq.\ (\ref{cp11}),
is shown in Fig.\ 4. Here
one observes that the time-like part of the hypersurface (dashed)
contributes less particles
than the space-like one (dash-dotted), but the slope is the same,
reflecting the fact that the temperature along the isothermal
freeze--out hypersurface is constant by definition.
The observed low--$k_{\bot}$ enhancement is due to
Bose--Einstein statistics.

For comparison, Fig.\ 5 shows the rapidity spectrum for isochronous
freeze--out, eq.\ (\ref{cp13}), with $\bar{T}=0.4\, T_0$.
The total spectrum (full line)
shows a double--peak structure with a local minimum at $y=0$ which
originates from the time dilation effect discussed in Section 2.
Different contributions from the ``time-like'' and ``space-like''
parts
as defined in Section 3.3 are shown separately (dashed, dash-dotted lines).
As is obvious from eq.\ (\ref{cp13}), there are
no negative contributions to the spectrum.

The corresponding transverse momentum spectrum, eq.\ (\ref{cp14}),
is presented in Fig.\ 6.
The ``space-like'' part of the fluid distribution has on the average
a higher temperature and thus its contribution to the transverse
momentum spectrum a smaller (negative) slope
than the corresponding spectrum from the ``time-like'' parts.
The total spectrum (full line) is nevertheless very
similar to the one in the isothermal freeze--out, Fig.\ 4,
because the {\em average\/} temperature at freeze--out is the same in
both scenarios.

We now study the dependence of the spectra on the freeze--out temperature.
Fig.\ 7 shows rapidity spectra for isothermal freeze--out
at $T_f=0.8,\, 0.6,\, 0.4,\, 0.2\, T_0$.
For the largest $T_f$--values
the rapidity spectrum peaks at finite $y$ while for
smaller $T_f$ also a peak at central rapidity $y=0$ develops.
Only for $T_f \ll T_0$ does this peak dominate the spectrum.

In the case of isochronous freeze--out the situation is quite
different. In Fig.\ 8, we show results for
$\bar{T}=0.8,\, 0.6,\, 0.4,\, 0.2\, T_0$.
For $\bar{T}=0.8\, T_0$, one observes a
rapidity distribution with a Gaussian-like shape, because in this case
freeze--out happens so early that the two maxima and the minimum at the origin
as seen in the temperature distribution (Fig.\ 2) are too close 
together to be resolved after thermal smearing (especially since in this case
the temperature is relatively high).
Also, for decreasing freeze--out temperatures there are considerable
differences to the case of isothermal freeze--out.
In particular, there is no maximum at central rapidity due to the
time dilation effect visible in the temperature distributions.

We now consider the variation of the transverse momentum spectrum
with temperature. In Fig.\ 9 we present the different
transverse momentum spectra for isothermal freeze--out and
for the same freeze--out temperatures as above.
As expected, lower freeze--out temperatures lead to steeper
distributions. The corresponding spectra for isochronous freeze--out
are shown in Fig.\ 10. Due to the above mentioned reasons, the 
results are very similar to those of isothermal freeze--out.

A Gaussian-like shape of the rapidity spectra is often claimed
to be a characteristic feature of the Landau expansion model
\cite{Landau,Srivastava}.
As is demonstrated in Fig.\ 7, this is not true in general and
especially not for realistic values of $T_f$.
A Gaussian-like spectrum emerges {\em only\/} if one neglects
the time-like contributions to the rapidity spectrum,
which, however, dominate the sum spectra at higher $T_f$.
To illustrate this point
we show in Fig.\ 11 the rapidity spectrum originating from
the space-like parts of the isothermal freeze--out hypersurface.
Note that in contrast to Fig.\ 7, {\em all rapidity spectra have
Gaussian-like shapes}. The neglected time-like contributions
are shown separately in Fig.\ 12.
We have normalized the spectra in Fig.\ 11 to their maximum value at $y=0$
in order to facilitate a comparison to the results of Srivastava et al.\
\cite{Srivastava} who employ this approximation.
Note also that their method to calculate the spectra
violates the conservation laws, since it does
not account for energy, momentum, entropy, or particle number
streaming through the time-like parts of the freeze--out hypersurface.

\section{Summary and Conclusions}

In this paper we have investigated freeze--out in a one--dimensional
hydrodynamic model for the expansion stage of heavy--ion collisions.
We have discussed two different freeze--out scenarios
used in the literature, namely instantaneous freeze--out
across an isothermal and an isochronous space--time hypersurface.

We have found that, provided $\bar{T}|_{t_f}=T_f$,
the transverse momentum spectra are very similar
in the two scenarios, but that there are noticeable differences
in the rapidity distributions. This can be explained as follows.
In our model, the fluid evolution is only one--dimensional and
variations in temperature and fluid velocity occur only along
the longitudinal direction. It is obvious that they consequently have
to vary along different hypersurfaces in space--time.
These differences are bound to reflect themselves
in the (longitudinal) rapidity distribution. On the other hand,
the transverse momentum spectra average over the longitudinal
degrees of freedom and thus are only influenced by the (average) temperature.
The fact that the latter is the same in both scenarios naturally explains
the observed similarity.
If one, however, selects a particular rapidity interval for the
calculation of the transverse momentum spectra, 
for instance midrapidity, the main contribution will come from
the space-like parts of the hypersurface 
at z=0, where the fluid temperature is different in the
two scenarios, cf. Fig. 1. The transverse momentum spectra will
reflect this difference, cf.\ Figs.\ 4 and 6.
Therefore, one should in general use the (more realistic) 
isothermal freeze--out scenario in three--dimensional calculations.
For central collisions with cylindrical symmetry, isothermal
freeze--out was already employed in the calculations of Ref.\ \cite{Ornik}.

We also demonstrated that an improper calculation of the particle
rapidity distribution, neglecting contributions from time-like parts
of the freeze--out hypersurface and thus violating the
conservation laws, leads to the well--known Gaussian-like shapes commonly
associated with the Landau expansion model. A consistent calculation, however,
produces double-- or triple--humped distributions.

Future studies should on the one hand focus on a more realistic description 
of the decoupling of particles from regions with finite space--time extent 
instead across an infinitely narrow space--time hypersurface.
On the other hand, the feedback of the decoupling process on the
evolution of the fluid has to be taken into account to obtain a fully  
self-consistent picture of freeze--out for the hydrodynamical description of
heavy--ion collisions.
\noindent
\\ ~~ \\
{\bf Acknowledgements} \\ ~~ \\
We thank Adrian Dumitru and Miklos Gyulassy
for fruitful and enlightening discussions, and Adrian Dumitru for a
careful reading of the manuscript.
J.A.M.\ thanks the Nuclear Theory Group at Columbia University for
its hospitality, where part of this work was done.

\newpage
\noindent
{\bf Figure Captions:}
\\ ~~ \\
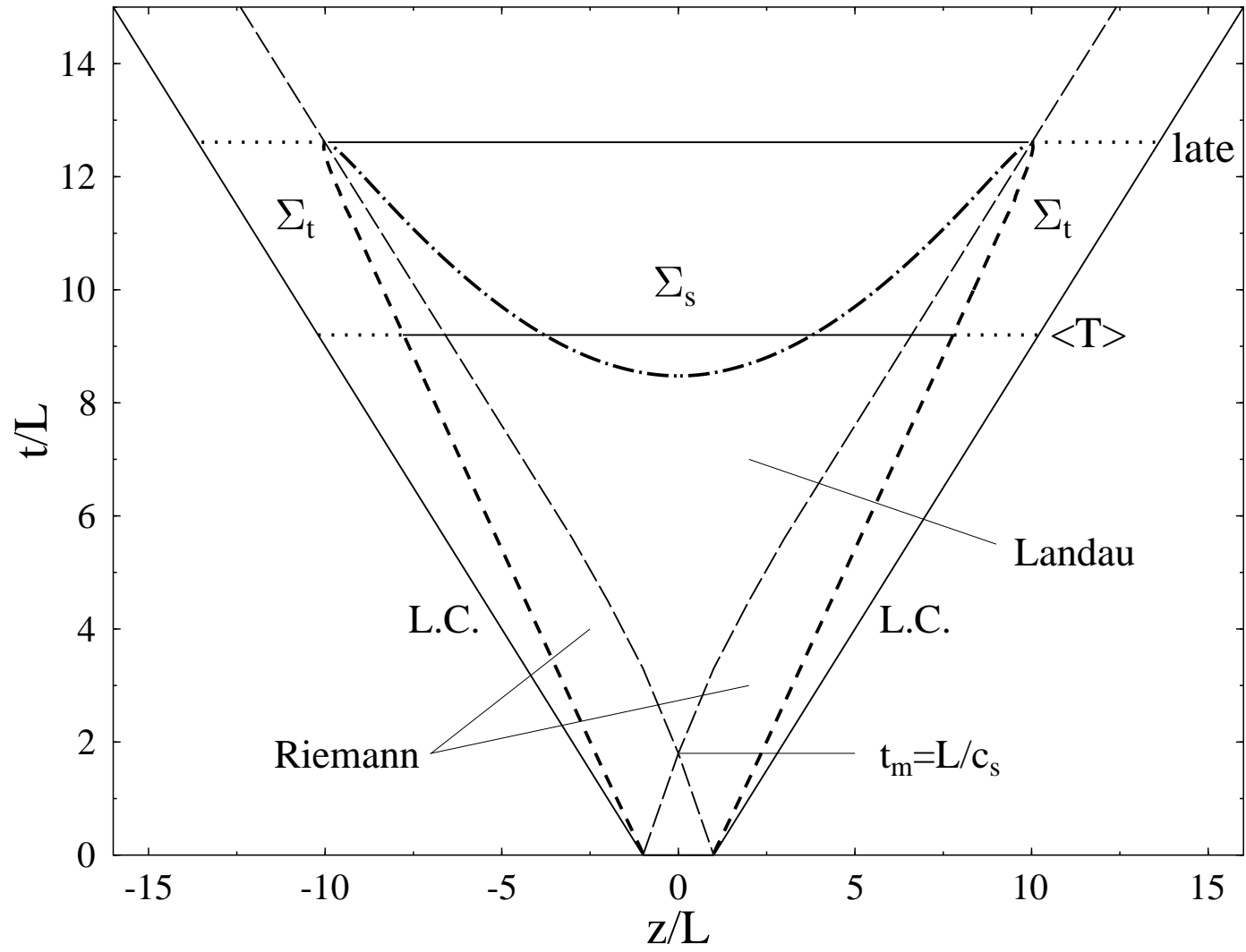
{\bf Fig.\ 1:}
Illustration of the different freeze--out scenarios. The expansion
has been calculated with the explicit SHASTA. As an example for the
isothermal freeze--out, we show the isotherm $T_f= 0.4\, T_0$.
It consists of two parts, namely the time-like $\Sigma_{t}$
(dashed), and the space-like $\Sigma_{s}$ (dash-dotted). 
For the isochronous freeze--out,
we show the (purely space-like) surfaces at the CM times when
(a) the average temperature $\bar{T}=0.4\, T_0$ is reached (lower line), and
(b) the temperature of the hottest cell has just dropped below $T_f$ 
(``late'' freeze--out, upper line). 
Also for the isochronous scenario we define so-called
``time-like'' (dotted) and ``space-like'' (full) contributions, for
an explanation see text. The thin dashed lines illustrate the boundaries
between the Riemann and the Landau region.
\\ ~~ \\
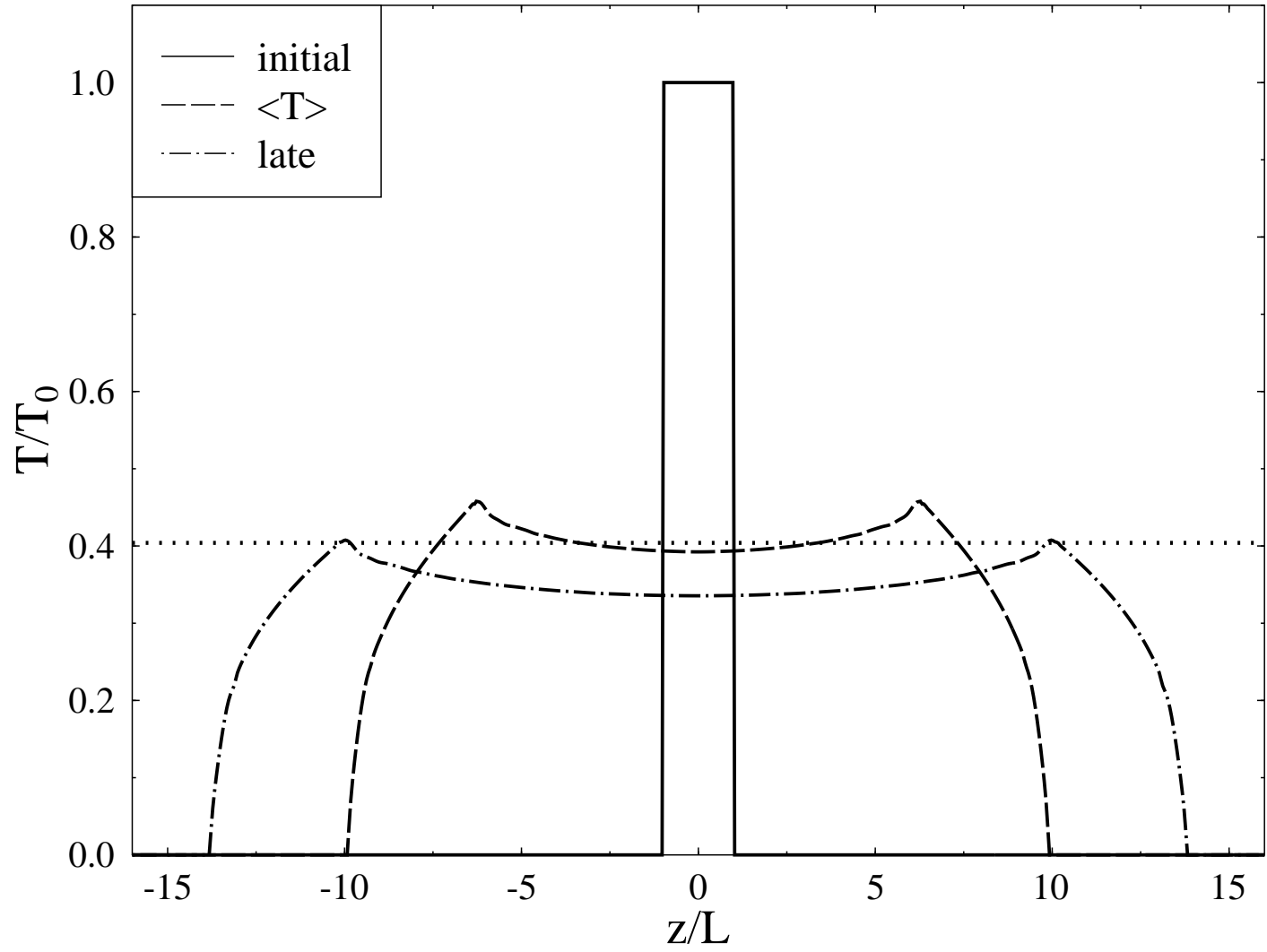
{\bf Fig.\ 2:}
Temperature profiles for the one-dimensional expansion, normalized
to the initial temperature $T_0$. Full line: initial profile.
Also shown are profiles at (a) $t \simeq 9.2\,\rm{L}$, when
the average temperature of the distribution is
$\bar{T}=0.4\, T_0$ (dashed), and (b) at $t \simeq 12.6\,\rm{L}$,
when the late freeze--out stage is reached (dash-dotted).
The profiles have been calculated with the explicit SHASTA and 
standard antidiffusion
\cite{RischkeBer}.
\\ ~~ \\
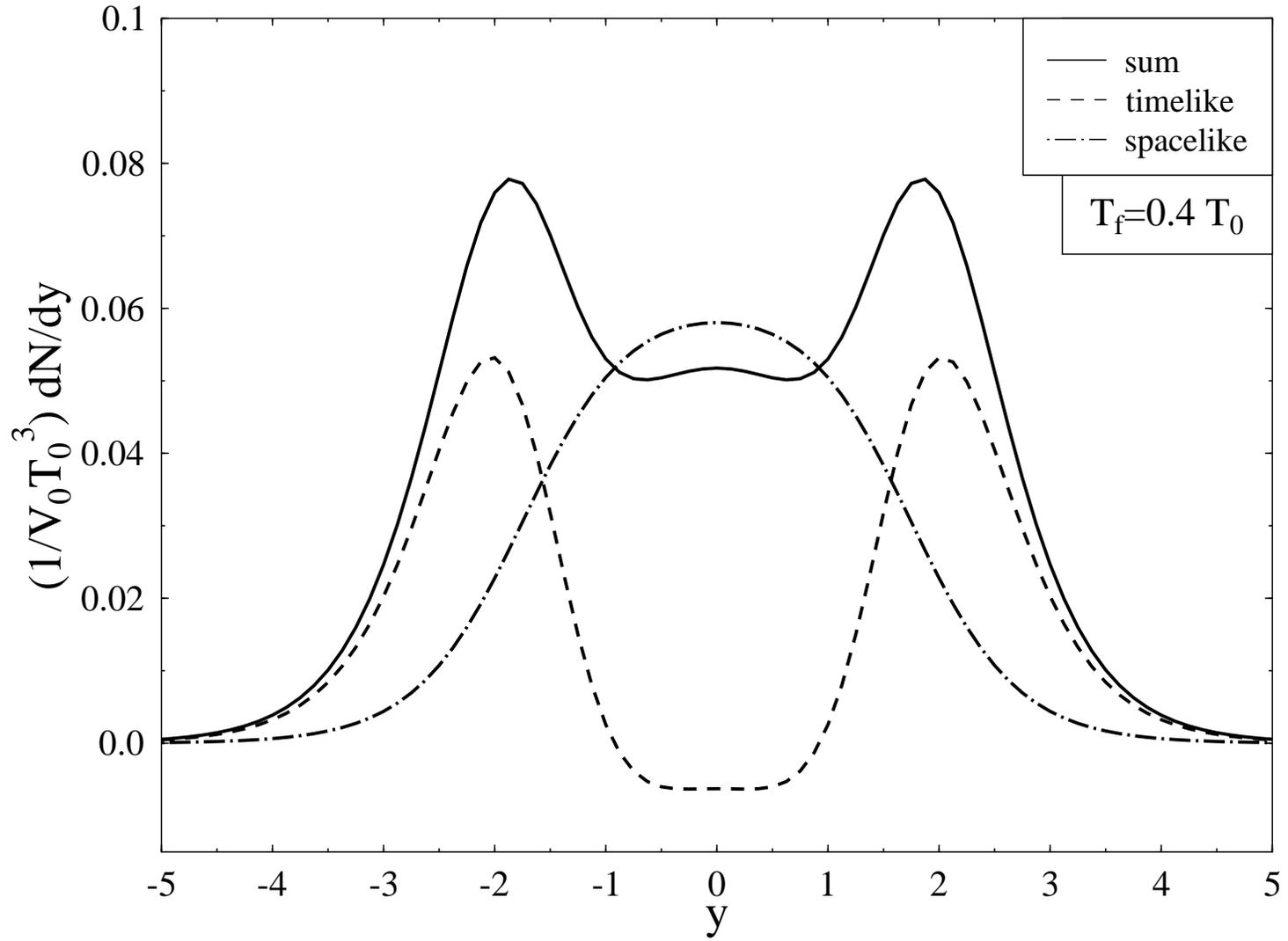
{\bf Fig.\ 3:}
Rapidity distribution (full line) for isothermal freeze--out
along the isotherm $T_f=0.4\, T_0$.
The distribution consists of contributions from
time-like (dashed) and space-like (dash-dotted)
parts of the hypersurface. 
$V_{0}=2LA$ is the initial volume of the expanding block.
\\ ~~ \\ 
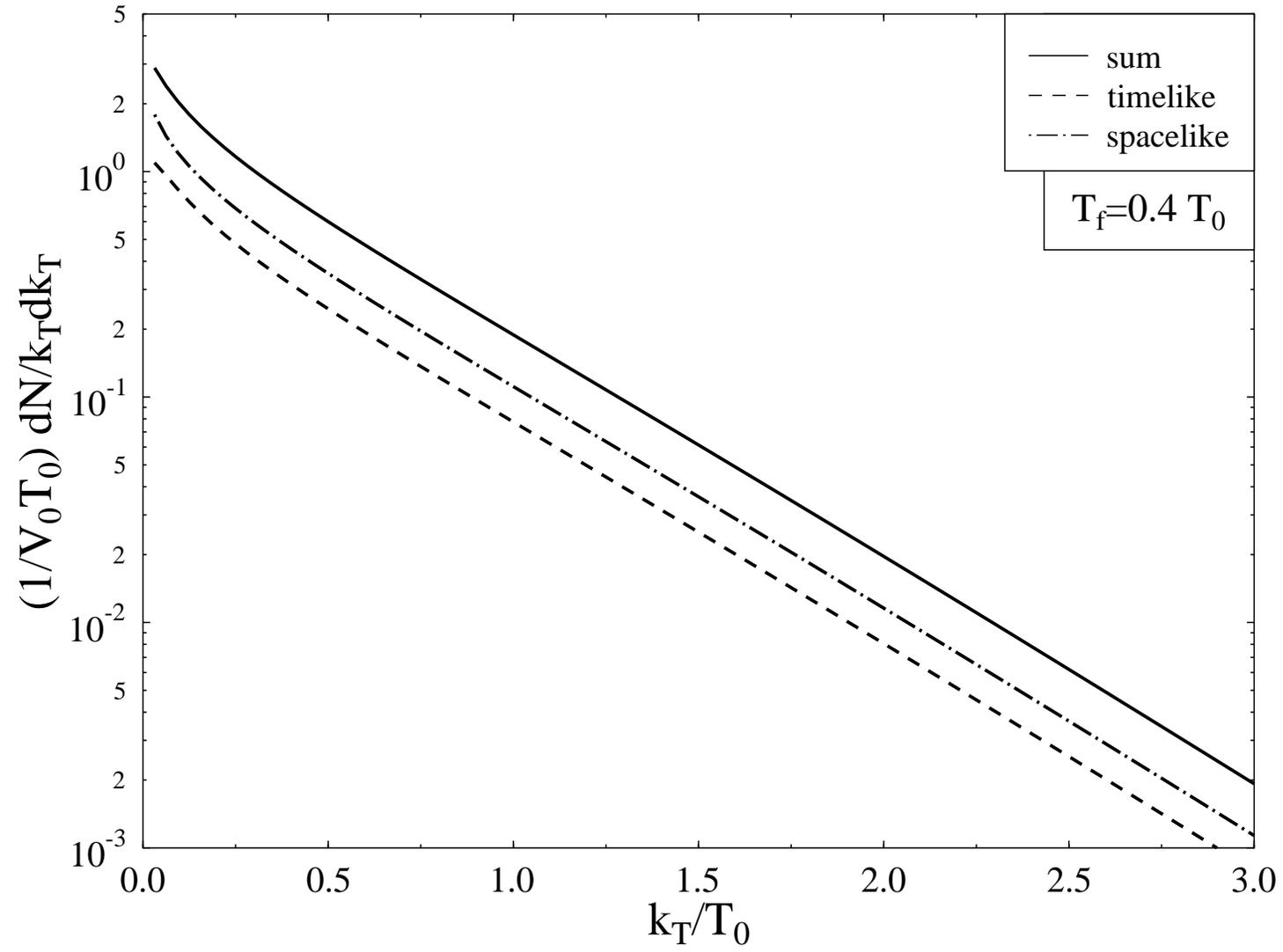
{\bf Fig.\ 4:}
Transverse momentum spectrum for isothermal freeze--out at
$T_f=0.4\, T_0$. Time-like (dashed) and space-like (dash-dotted)
contributions are shown separately.
\\ ~~ \\
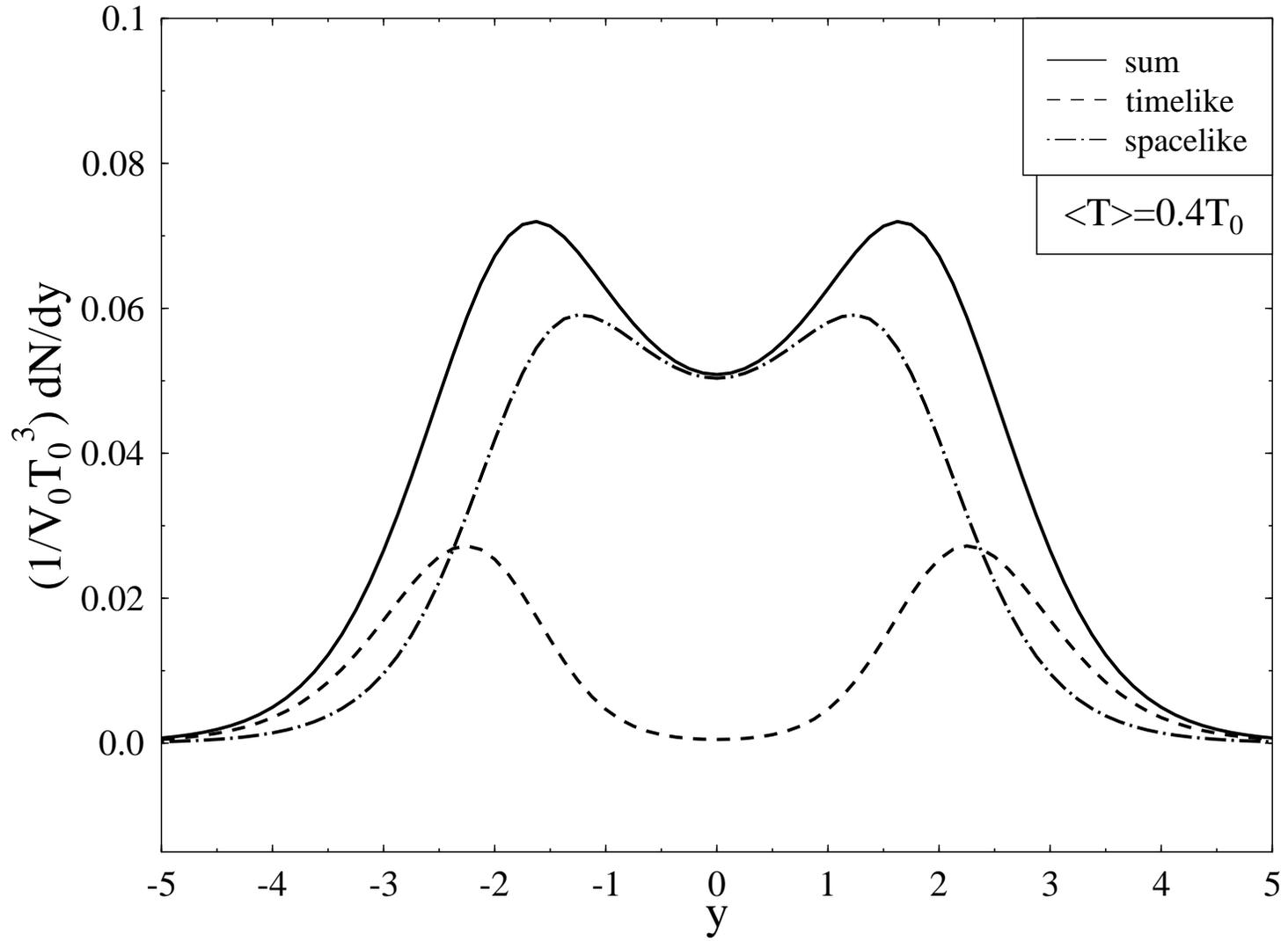
{\bf Fig.\ 5:}
Rapidity distribution (full line) for isochronous freeze--out
with $\bar{T}=0.4\, T_0$.
The distribution consists of contributions from the
``time-like'' (dashed)
and ``space-like'' (dash-dotted) parts of the hypersurface.
\\ ~~ \\
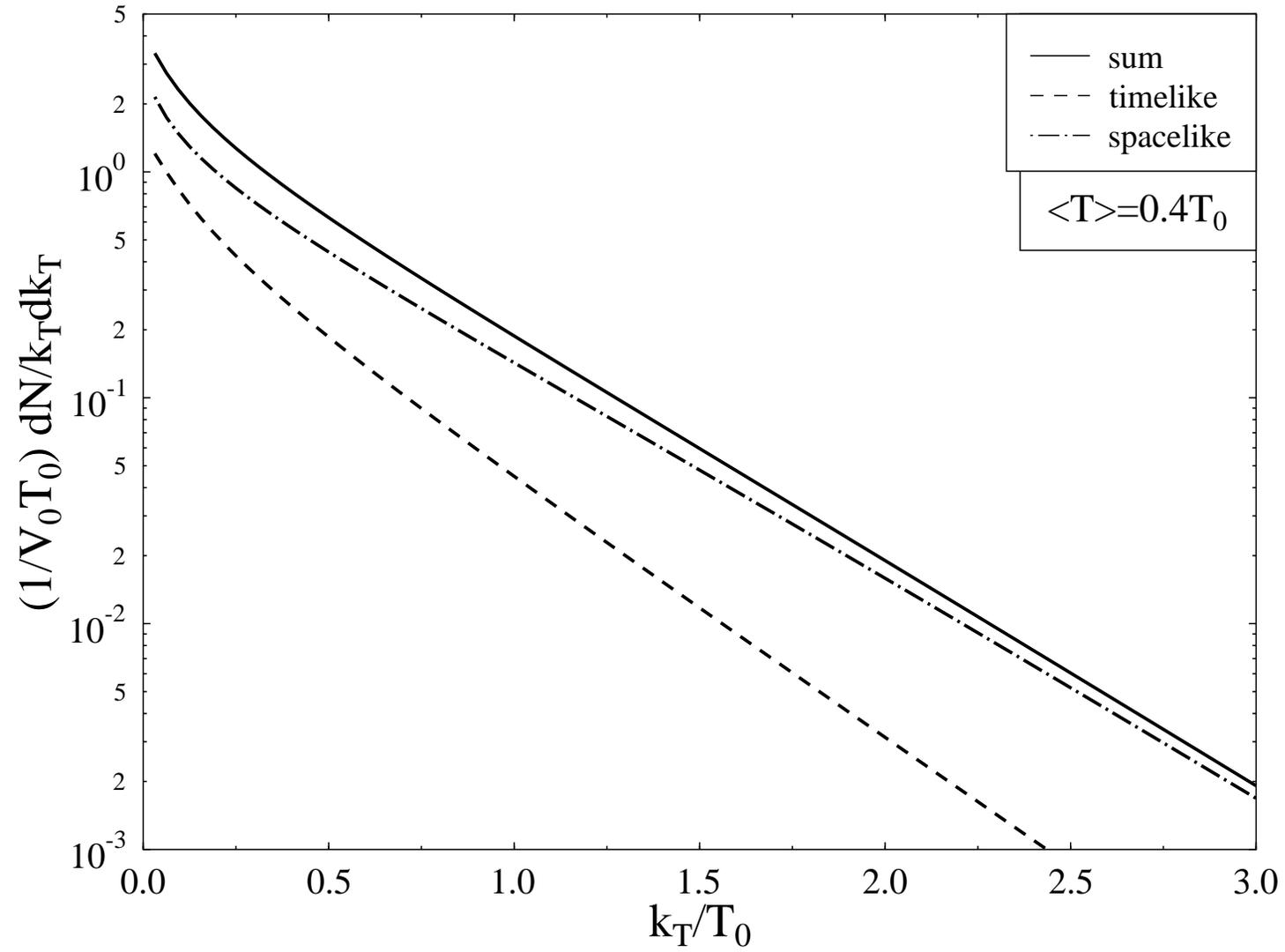
{\bf Fig.\ 6:}
Transverse momentum spectrum for isochronous freeze--out with
$\bar{T}=0.4\, T_0$. ``Time-like'' (dashed)
and ``space-like'' (dash-dotted) contributions are shown
separately.
\\ ~~ \\
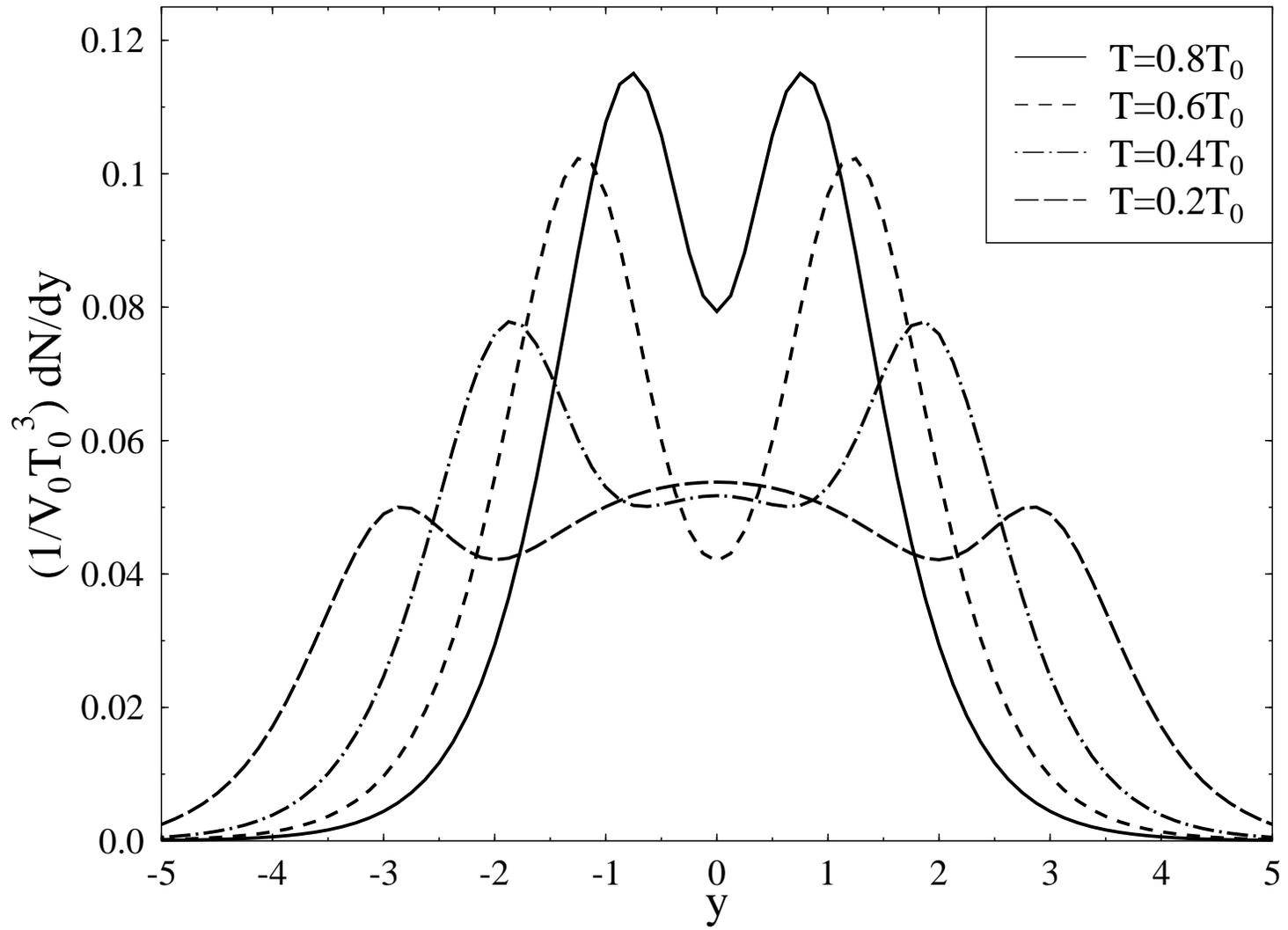
{\bf Fig.\ 7:}
Rapidity distributions for isothermal freeze--out
along the isotherms $T_f/T_0=0.8,\, 0.6,\, 0.4,\, 0.2$.
Only the total distributions are shown.
\\ ~~ \\
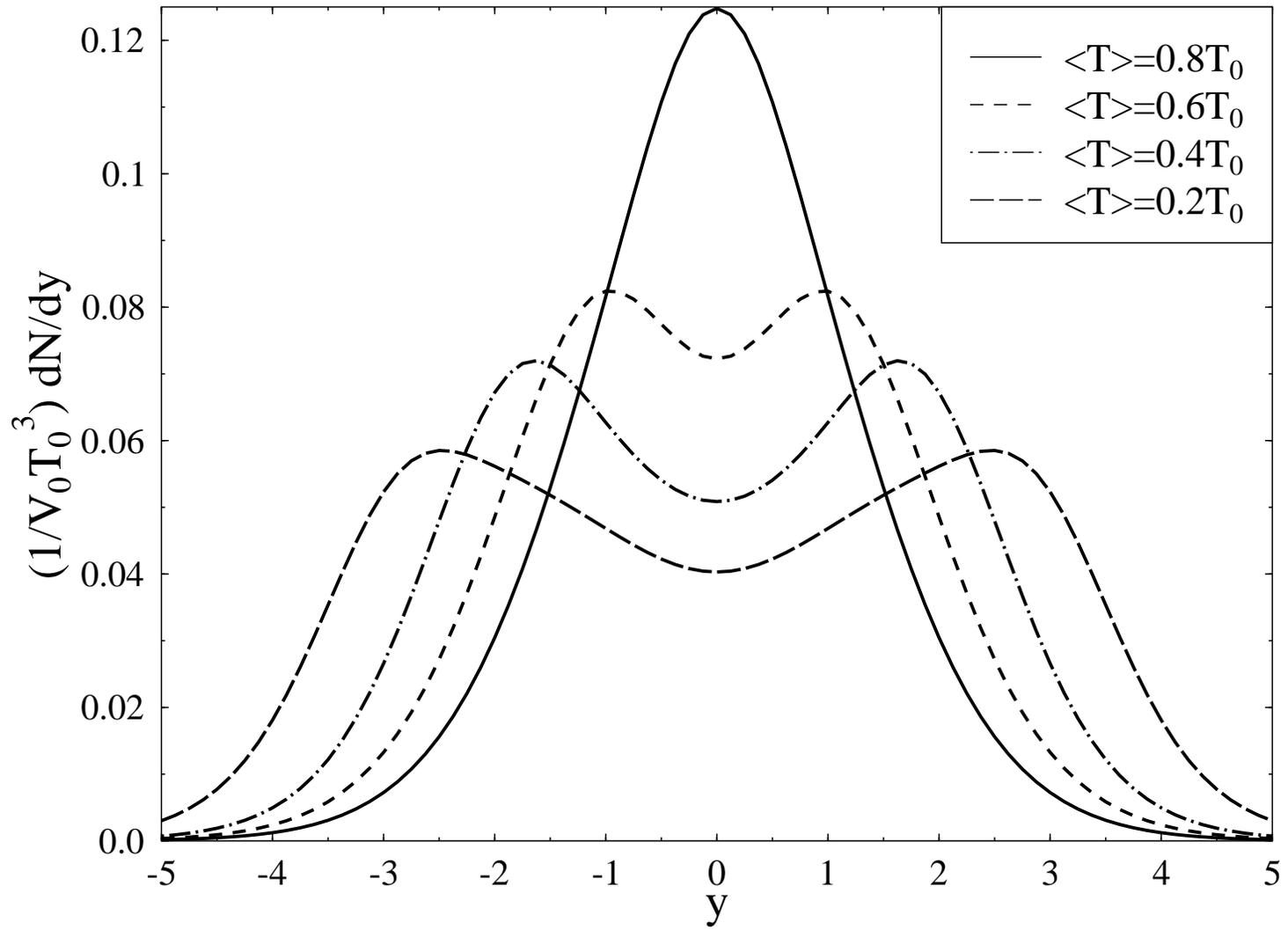
{\bf Fig.\ 8:}
Rapidity distributions for isochronous freeze--out
at average temperatures $\bar{T}/T_0=0.8,\, 0.6,\, 0.4,\, 0.2$.
Only the total distributions are shown.
\\ ~~ \\
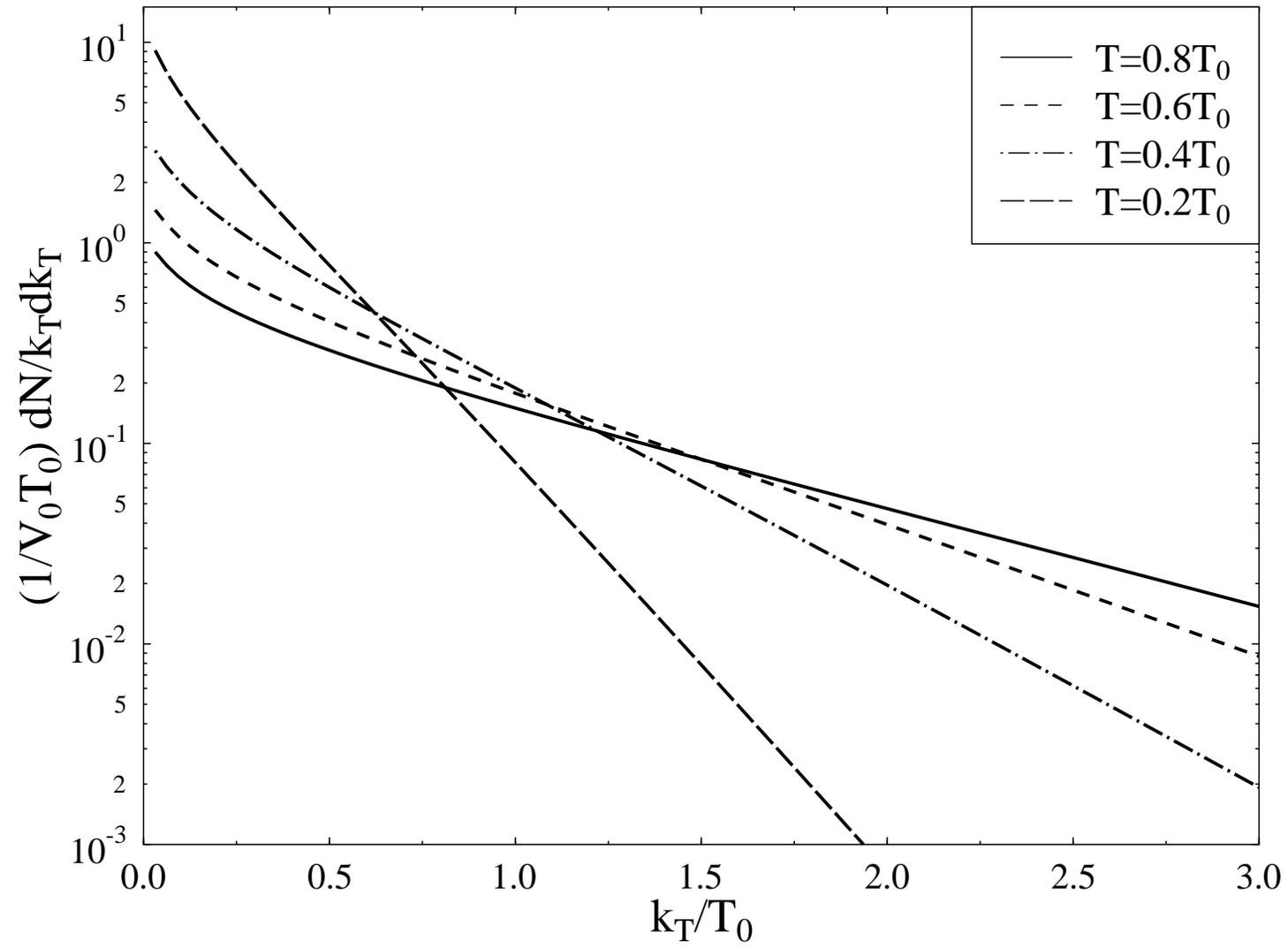
{\bf Fig.\ 9:}
Transverse momentum spectra for isothermal freeze--out
along the isotherms $T_f/T_0=0.8,\, 0.6,\, 0.4,\, 0.2$.
Only the total spectra are shown.
\\ ~~ \\
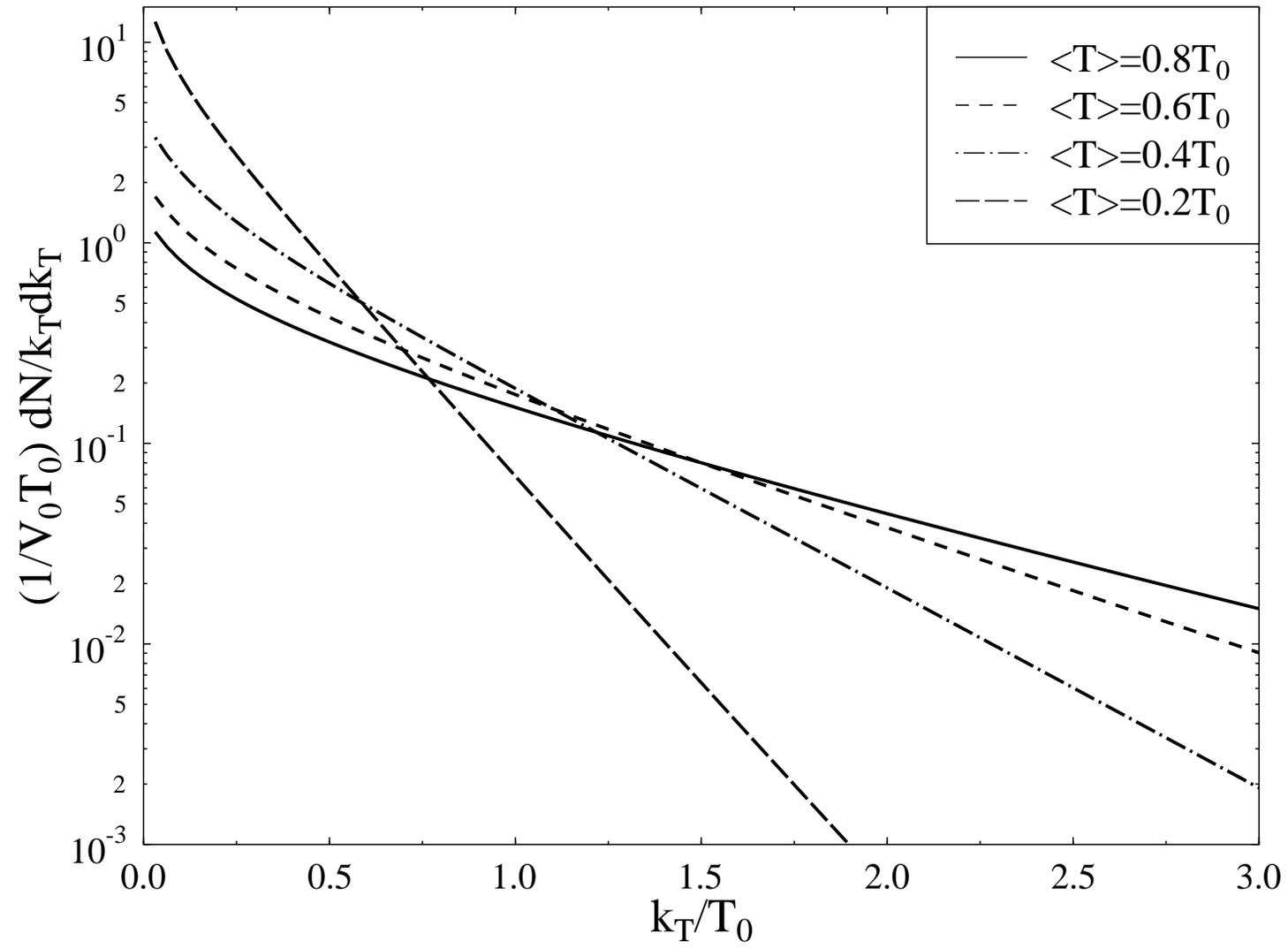
{\bf Fig.\ 10:}
Transverse momentum spectra for isochronous freeze--out
at average temperatures $\bar{T}/T_0=0.8,\, 0.6,\, 0.4,\, 0.2$.
Only the total spectra are shown.
\\ ~~ \\
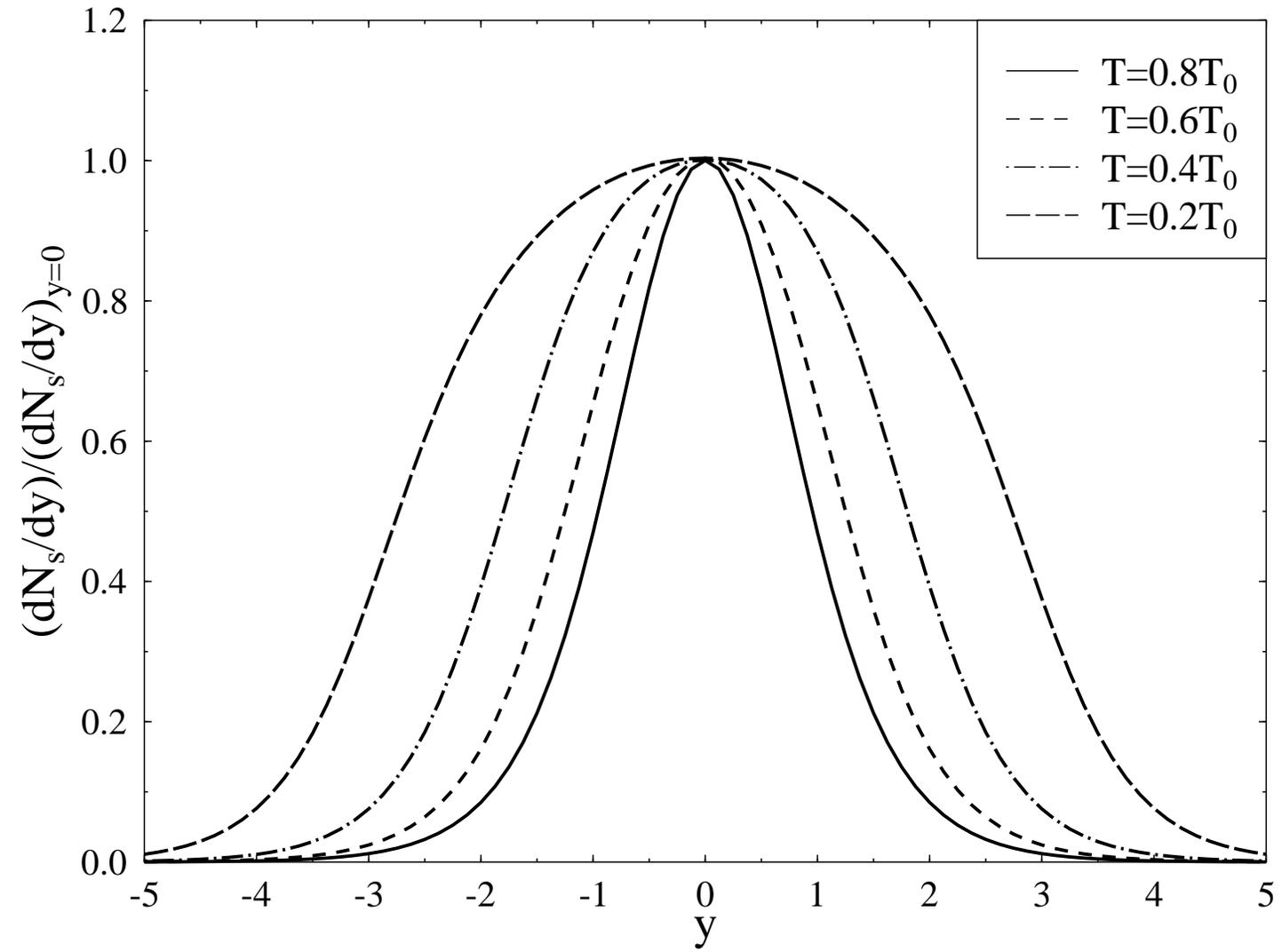
{\bf Fig.\ 11:}
Rapidity distributions along the space-like part of the
hypersurfaces for isothermal freeze--out
with temperatures $T_f/T_0=0.8,\, 0.6,\, 0.4,\, 0.2$.
The distributions have been normalized to their maximum value.
The time-like parts of the hypersurfaces are neglected.
\\ ~~ \\
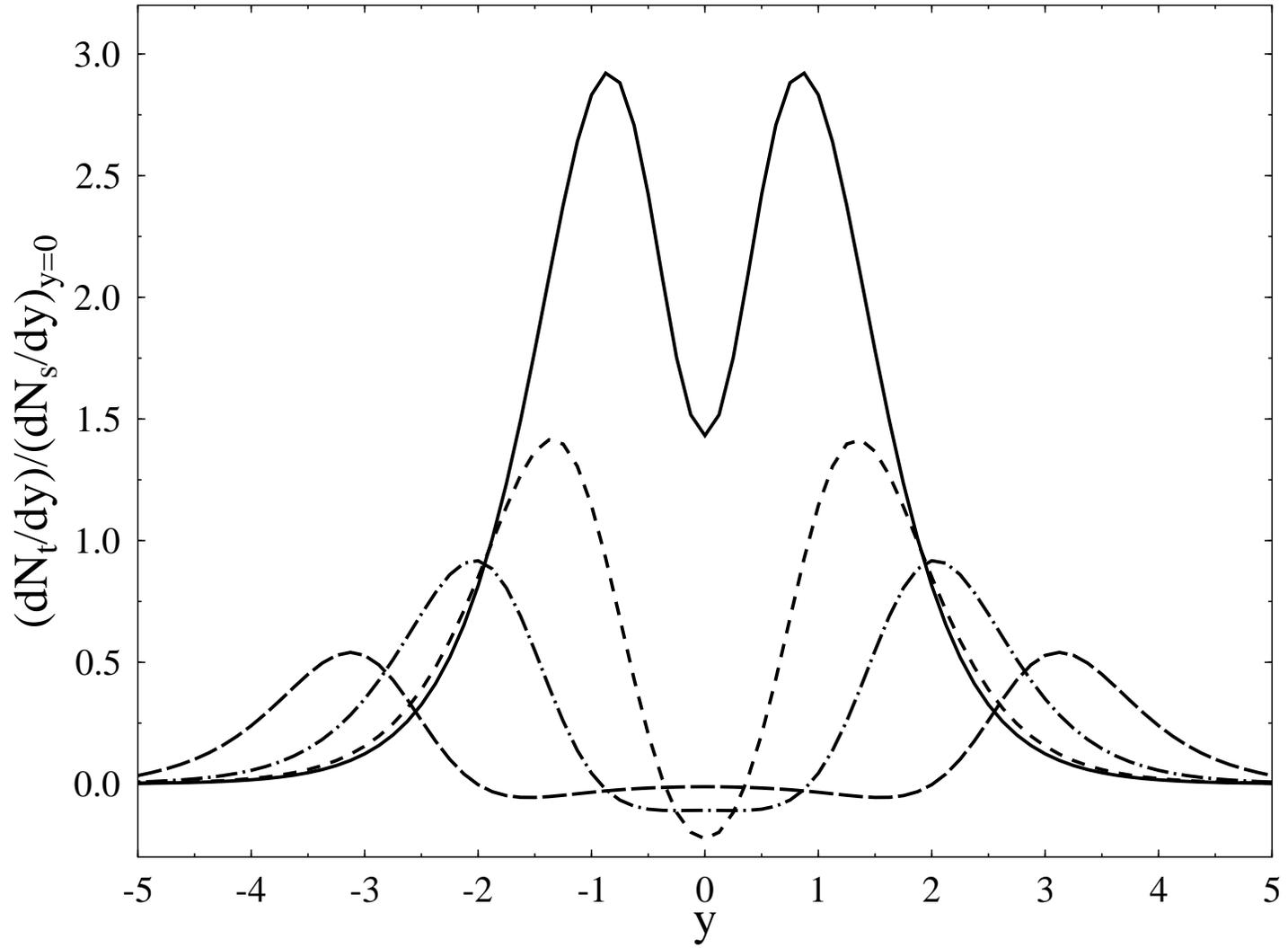
{\bf Fig.\ 12:}
The time-like contributions that have been neglected in
Fig.\ 11. 
The distributions are normalized to the corresponding maxima in Fig.\ 11.
\end{document}